\documentclass[aps,prd,twocolumn,floatfix,nofootinbib,showpacs,superscriptaddress,tightenlines]{revtex4}

\usepackage{amsmath}
\usepackage{lipsum}
\usepackage{amssymb}
\usepackage{amsthm}
\usepackage{bbold}
\usepackage{dcolumn}
\usepackage{epsfig}
\usepackage{graphics}
\usepackage{graphicx}
\usepackage{longtable}
\usepackage{color}
\usepackage{bm}
\usepackage{xspace}
\usepackage{cancel}

\definecolor{darkgreen}{rgb}{0,0.5,0}
\definecolor{purple}{rgb}{0.5,0,0.5}
\definecolor{nblue}{rgb}{0.0,0.0,0.50}
\definecolor{scarlet}{rgb}{1.0,0.2,0}

\usepackage[colorlinks=true, pdfstartview=FitV, linkcolor=purple, citecolor= purple, urlcolor=blue]{hyperref}

\newcommand{\be}{\begin{equation}}
\newcommand{\ee}{\end{equation}}
\newcommand{\bea}{\begin{eqnarray}}
\newcommand{\eea}{\end{eqnarray}}
\newcommand{\nn}{\nonumber}
\newcommand{\beas}{\begin{eqnarray*}}
\newcommand{\eeas}{\end{eqnarray*}}

\newcommand{\half}{{\textstyle\frac{1}{2}}}
\newcommand{\kslash}{\not\!{k}}
\newcommand{\pslash}{\not\!{p}}
\newcommand{\qslash}{\not\!{q}}

\begin{document}
\title{Quark-gluon Vertex: A Perturbation Theory Primer and Beyond}

\author{R. Bermudez}
\affiliation{Departamento de Investigaci\'on en F\'{\i}sica,
Universidad de Sonora, Boulevard Luis Encinas J. y Rosales,
Colonia Centro, Hermosillo, Sonora 83000, M\'exico}

\author{L. Albino}
\affiliation{Instituto de F\'isica y Matem\'aticas, Universidad
Michoacana de San Nicol\'as de Hidalgo, Edificio C-3, Ciudad
Universitaria, Morelia, Michoac\'an 58040, M\'exico}

\author{L.X. Guti\'errez-Guerrero}
\affiliation{CONACyT-Mesoamerican Centre for Theoretical Physics,
Universidad Aut\'onoma de Chiapas, Carretera Zapata Km. 4, Real
del Bosque (Ter\'an), Tuxtla Guti\'errez 29040, Chiapas, M\'exico}

\author{M. E. Tejeda-Yeomans}
\affiliation{Departamento de F\'{\i}sica, Universidad de Sonora,
Boulevard Luis Encinas J. y Rosales, Colonia Centro, Hermosillo,
Sonora 83000, M\'exico}

\author{A. Bashir}
\affiliation{Instituto de F\'isica y Matem\'aticas, Universidad
Michoacana de San Nicol\'as de Hidalgo, Edificio C-3, Ciudad
Universitaria, Morelia, Michoac\'an 58040, M\'exico}

\begin{abstract}

There has been growing evidence that the infrared enhancement of
the form factors defining the full quark-gluon vertex plays an
important role in realizing a dynamical breakdown of chiral
symmetry in quantum chromodynamics, leading to the observed
spectrum and properties of hadrons. Both the lattice and the
Schwinger-Dyson communities have begun to calculate these form
factors in various kinematical regimes of momenta involved. A
natural consistency check for these studies is that they should
match onto the perturbative predictions in the ultraviolet, where
non-perturbative effects mellow down. In this article, we carry
out a numerical analysis of the one-loop result for all the form
factors of the quark-gluon vertex. Interestingly, even the
one-loop results qualitatively encode most of the infrared
enhancement features expected of their non-perturbative counter
parts. We analyze various kinematical configurations of momenta:
symmetric, on-shell and asymptotic. The on-shell limit enables us
to compute anomalous chromomagnetic moment of quarks. The
asymptotic results have implications for the multiplicative
renormalizability of the quark propagator and its connection with
the Landau-Khalatnikov-Fradkin transformations, allowing us to
analyze and compare various {\em Ans$\ddot{a}$tze} proposed so
far.

\end{abstract}


\maketitle

%%%%%%%%%%%%%%%%%%%%%%%%%%%%%%%%%%%%%%%%%%%%%%%%%%%%%%%%%%%%%%%%%
%%%%%%%%%%%%%%%%%%%%%%%%%%%%%%%%%%%%%%%%%%%%%%%%%%%%%%%%%%%%%%%%%
\section{Introduction}
%%%%%%%%%%%%%%%%%%%%%%%%%%%%%%%%%%%%%%%%%%%%%%%%%%%%%%%%%%%%%%%%%
%%%%%%%%%%%%%%%%%%%%%%%%%%%%%%%%%%%%%%%%%%%%%%%%%%%%%%%%%%%%%%%%%

Non-perturbative study of the Schwinger-Dyson equation (SDE) for
the quark propagator has suggested infrared enhancement of the
running quark mass function ${\cal M}(p^2)$ through the dynamical
breakdown of chiral
symmetry,~\cite{Maris:1997tm,Bhagwat:2006tu,Williams:2006vva}.
Lattice studies have provided its
confirmation,~\cite{Bowman:2002bm,Parappilly:2005ei,Furui:2006ks}.
It is also well-known that the corresponding quark propagator
breaches the axiom of reflection positivity and hence corresponds
to a confined excitation,~\cite{Bashir:2012fs,Fischer:2006ub}.
However, it is important to note that the analytic structure of
the quark propagator depends strongly on the details of the
structure of the quark-gluon vertex, which makes the study of the
latter all the more important,~\cite{Fischer:2006ub}. Whereas in
the infrared, i.e., ${\cal M}(p^2)|_{p^2 \rightarrow 0}$, the
quark mass function obtains a constituent-like value of about
$300-500$ MeV, its perturbative limit is reproduced correctly in
the ultraviolet domain. This running of the mass function has
innumerable observable consequences in hadron physics, see for
example recent reviews,~\cite{Bashir:2012fs,Aznauryan:2012ba}.

The quark propagator is intimately linked with the corresponding
behavior of the gluon propagator and the quark-gluon vertex
through the relevant SDEs as well as the symmetry relations of
quantum chromodynamics (QCD), namely the Slavnov-Taylor identities
(STIs),~\cite{Slavnov:1972fg,Taylor:1971ff}, the transverse
Takahashi identities
(TTIs),~\cite{Takahashi:1985yz,Kondo:1996xn,He:2000we},
 and the generalized
Landau-Khalatnikov-Fradkin transformations
(LKFTs),~\cite{Aslam:2015nia}. Therefore, a knowledge of the gluon
propagator and the quark-gluon vertex is vital to study their
impact on the quark propagator and the dynamical mass generation.

In the last decade or so, valuable conclusions have been arrived
at regarding the gluon propagator and a gluonic mass scale of
about $(2-4) \Lambda_{\rm QCD}$ associated with it in the
infrared. The SDEs prediction for the massive gluon
solution,~\cite{Aguilar:2004sw}, has also been confirmed in modern
lattice
studies,~\cite{Cucchieri:2007md,Bogolubsky:2007ud,Cucchieri:2010xr,Bogolubsky:2009dc,Oliveira:2009eh},
which support a finite but infrared enhanced scalar form factor of
the gluon propagator, the so called decoupling solution. It is
also in agreement with subsequent SDE
results,~\cite{Aguilar:2008xm,Boucaud:2008ky,Pennington:2011xs},
exact renormalization group (RG) equations,~\cite{Fischer:2008uz},
refined Gribov-Zwanziger
formalism,~\cite{Dudal:2007cw,Dudal:2008sp,Dudal:2010tf,Cucchieri:2011ig},
and the earlier suggestion of Cornwall,~\cite{Cornwall:1981zr}.
Even if one includes the effect of dynamical
quarks,~\cite{Bowman:2007du,Ayala:2012pb,Aguilar:2012rz}, the
qualitative behavior of the gluon propagator remains unaltered.
The screening effect of the increasing number of flavors is
reflected in the reduction in infrared strength of the gluon
propagator. Moreover, its feedback into the quark
propagator,~\cite{Bashir:2013zha}, is similar to what is observed
in quantum electrodynamics
(QED),~\cite{Bashir:2011ij,Akram:2012jq}. Interestingly, the
analytic properties of the gluon propagator do not permit it to
propagate freely,~\cite{Strauss:2012dg}. Again, a transition of
the associated scalar form factor to the perturbative limit of
$p^2 \gg \Lambda^2_{\rm QCD}$ is faithfully achieved. All the
above findings thus conform with the fact that individual quarks
and gluons are confined within color singlet hadrons.

In addition to the gluon propagator, the quark-gluon vertex also
feeds into the SDE of the quark propagator. Attempts have been
initiated in lattice QCD to compute the form factors of the
quark-gluon vertex for some symmetric kinematical configurations
of momenta
involved,~\cite{Skullerud:2002ge,Skullerud:2003qu,Skullerud:2004gp,Kizilersu:2006et}.
The SDEs can access any kinematical configuration of the external
momenta with the same amount of effort. Studies have been carried
out to see if the SDE truncations agree with the infrared
enhancement of the form factors reported in the lattice
computation,~\cite{Bhagwat:2004kj,Bhagwat:2004hn,LlanesEstrada:2004jz}.
A satisfactory agreement between the lattice and the SDE results
reassures that these two approaches are complementary.

The theoretical and phenomenological implications of different
form factors, defining the quark-gluon vertex, can hardly be
over-emphasized. For example, the mass splitting between the
parity partners in low lying mesons, such as $(\pi,\sigma)$ and
$(\rho,a_1)$, can only be explained through incorporating the form
factors proportional to the anomalous
chromomagnetic/electromagnetic vector structure $q_{\mu}
\sigma^{\mu \nu}$,~\cite{Chang:2010hb}. The associated corrections
cancel for the pseudoscalar and vector mesons but add in the
scalar and axial vector channels,~\cite{Chang:2009zb}, hence
solving a long standing puzzle. On the other hand, the choice of
the quark-gluon vertex (and quark-photon vertex, as the hadrons
are probed through photons) is also critically important in
studying the form factors of
mesons,~\cite{Raya:2015gva,Raya:2016yuj},  and
baryons,~\cite{Cloet:2013gva}.

Just like the quark mass function and the gluon propagator, the
form factors of the vertex should reduce to their perturbative
Feynman expansion in the weak coupling limit. Recall that a
truncation of the complete set of SDEs, which maintains gauge
invariance and multiplicative renormalizability (MR) at every
level of approximation, is perturbation theory. In QED, this fact
has long been used to impose constraints on the {\em Ansatz}
proposed for the fermion-boson vertex, see for
example~\cite{Curtis:1990zs,Bashir:1994az,Bashir:1997qt,Bashir:1999bd,Bashir:2007qq,Kizilersu:2009kg,Bashir:2011vg,Bashir:2011dp,Fernandez-Rangel:2016zac,Binosi:2016wcx}.
There are several one-loop results, available over the past three
decades, which facilitate this
task,~\cite{Curtis:1990zs,Ball:1980ay,Kizilersu:1995iz}. In this
article, we shall employ the one-loop perturbative calculation of
the quark-gluon vertex,~\cite{Davydychev:2000rt}, to deduce a
series of analytical and numerical requirements which any
non-perturbative construction of this vertex must comply with in
the weak coupling regime. Once it is achieved, the corresponding
truncation scheme encodes a more reliable transition from infrared
to the ultraviolet behavior of QCD.

This article is organized as follows: In section~\ref{sec:decomp}
we present the general considerations regarding the construction
of a physically meaningful and reliable quark-gluon vertex
\textit{Ansatz}. In section~\ref{sec:SymLimit}, analytical and
numerical computations of the vertex form factors for the
so-called symmetric limit (equal incoming, outgoing quark and
gluon momenta squared) are presented. In
section~\ref{sec:AsympLimit}, we provide analytical results for
the kinematical regime, where the momentum squared in one of the
quark legs is much larger than in the other, namely the asymptotic
limit. In section~\ref{sec:OnShellLimit}, we discuss the
physically relevant anomalous chromomagnetic moment of quarks in
the on-shell limit where it is historically defined. Finally, in
section~\ref{sec:Conclusions}, we present our conclusions and
final remarks.

%%%%%%%%%%%%%%%%%%%%%%%%%%%%%%%%%%%%%%%%%%%%%%%%%%%%%%%%%%%%%%%%%
%%%%%%%%%%%%%%%%%%%%%%%%%%%%%%%%%%%%%%%%%%%%%%%%%%%%%%%%%%%%%%%%%
\section{The Quark-Gluon Vertex - General Considerations}
\label{sec:decomp}
%%%%%%%%%%%%%%%%%%%%%%%%%%%%%%%%%%%%%%%%%%%%%%%%%%%%%%%%%%%%%%%%%
%%%%%%%%%%%%%%%%%%%%%%%%%%%%%%%%%%%%%%%%%%%%%%%%%%%%%%%%%%%%%%%%%

The quark-gluon vertex plays a fundamental role in perturbation
theory and in the non-perturbative treatment of QCD and hadron
physics. Therefore, we set out to study it in detail.

We start by expanding out this vertex in a tensor decomposition
dictated by a necessary constraint of gauge invariance, i.e., the
STI. We follow the procedure outlined in QED
in~\cite{Ball:1980ay,Kizilersu:1995iz,Bashir:2007qq} and adopted
for QCD in~\cite{Davydychev:2000rt}. The
STI,~\cite{Slavnov:1972fg}, which relates the quark-gluon vertex
$\Gamma_\mu\equiv\Gamma_\mu (p,k,q)$ with the quark propagator,
reads as follows:
\begin{equation}
q^\mu \Gamma_\mu=G(q^2)[ \overline{H}(k,p,q) S^{-1}(k) - S^{-1}(p)
H(p,k,q) ] \,, \label{WTSI-full}
\end{equation}
where $q=k-p$ and $G(q^2)$ is the scalar function associated with
the ghost propagator. The function $H$, and its ``conjugated''
function $\overline{H}$, are related to the auxiliary non-trivial
vertices involving the complete four-point quark-quark-ghost-ghost
vertex. $k$ and $p$ are the incoming and outgoing quark momenta,
respectively, while $q$ is the outgoing gluon momentum. Moreover,
$S(k)$ is the full quark propagator, defined as
\begin{equation}
S(k) = \frac{F(k^{2})}{ \not\!{k} - {\cal{M}} (k^{2}) } \,,
\end{equation}
where $F(k^2)$ is the so-called wave function renormalization, and
${\cal{M}}(k^2)$ is the running quark mass function. At the tree
level, $F(k^2)=1$ and ${\cal{M}} (k^{2})=m$, the current quark
mass. Similarly, we define the tree level gluon propagator as:
 \bea
 \Delta_{\mu \nu}(q^2) = -i \frac{1}{q^2} \left[ g_{\mu \nu} - \xi
 \frac{q_{\mu} q_{\nu}}{q^2} \right] \;,
 \eea
where $\xi$ is the covariant gauge parameter. $\xi=0$ is the
Feynman gauge while $\xi=1$ corresponds to the Landau gauge.

Finally, the vector $\Gamma_\mu (p,k,q)$ stands for the
fully-dressed three-point quark-gluon vertex. Below we enlist the
conditions which constrain its construction:

\begin{itemize}

 \item It must satisfy the STI. This implies that the requirement of
gauge invariance fixes the longitudinal part of the quark-gluon
vertex.

 \item The transverse part is constrained by the requirement of
the MR of the massless quark propagator, the LKFTs and the TTIs.

 \item The tensor decomposition selected guarantees that every coefficient
$\tau_i$ should be free of kinematic singularities when
$k^2\rightarrow p^2$ at the one-loop level in arbitrary covariant
gauge and
dimensions,~\cite{Ball:1980ay,Curtis:1990zs,Davydychev:2000rt,Bashir:1999bd}.
We expect it to be true non-perturbatively too because the only
singularities which arise are due to good dynamical reasons such
as the mass poles for physical particles.

 \item The vertex must transform under the charge conjugation ($C$), parity ($P$), and time
 reversal ($T$) operations just as the bare vertex.

 \item It should reduce to its perturbation theory Feynman
expansion in the limit of weak coupling. Note that a truncation of
the complete set of SDEs, which maintains gauge invariance and MR
of a gauge theory at every level of approximation, is perturbation
theory. Therefore, physically meaningful solutions of the SDEs
must agree with perturbative results in the weak coupling regime.
In this article, we use one-loop perturbative calculation of the
quark-gluon vertex,~\cite{Davydychev:2000rt}, as a guiding
principle to impose tight constraints on the quark-gluon vertex.

\end{itemize}

Starting from the STI, Eq.~(\ref{WTSI-full}), we can decompose the
vertex as a sum of longitudinal and transverse
components,~\cite{Ball:1980ay}:
\begin{equation} \Gamma_{\mu} (p,k,q) =
\Gamma_{\mu}^{L} (p,k,q)+\Gamma_{\mu}^{T} (p,k,q) \,,\label{F-LT}
\end{equation}
where the longitudinal part $\Gamma_{\mu}^{L} (p,k,q)$ alone
satisfies the STI,~Eq.~(\ref{WTSI-full}), and the transverse part,
$\Gamma_{\mu}^{T} (p,k,q)$, is naturally constrained by the
conditions
 \bea
 q^{\mu} \Gamma_{\mu}^{T} (p,k,q)=0 \,, \quad \Gamma_{\mu}^{T}
 (p,p,0)= 0 \,.
 \eea
 This decomposition ensures
that all ultraviolet (UV) divergences are encoded in the
longitudinal component, which in turn is expressed as
 \bea \label{L-Vertex}
\Gamma_{\mu}^{L}(p,k,q) = \sum_{i=1}^{4}
\lambda_{i}(p^{2},k^{2},q^{2}) L^i_{\mu}(p,k) \,.
 \eea
The longitudinal tensor basis acquires the form~:
 \bea \label{L_i}
L^1_{\mu} &=& \gamma_{\mu} \,,
\nn \\
L^2_{\mu}(p,k) &=& (\pslash \, +\kslash)(p+k)_{\mu} \,,
\nn \\
L^3_{\mu}(p,k) &=& -(p+k)_{\mu} \,,
\nn \\
L^4_{\mu}(p,k) &=& -\sigma_{\mu \nu}\, (p+k)^{\nu} \,,
 \eea
where $\sigma_{\mu\nu}=\frac{1}{2}[\gamma_\mu,\gamma_\nu]$. The
longitudinal form factors $\lambda_i$ of
Eq.~(\ref{L-Vertex})\footnote{In perturbation theory, we will
express $\lambda_1 \rightarrow 1+ \lambda_1$ for the sake of
convenience, separating out the tree-level factor of 1.} are
expressed in terms of the scalar functions associated with the
quark, gluon and ghost propagators and the four-point quark-ghost
vertices,~\cite{Aguilar:2010cn}. On the other hand, the UV-finite
transverse component is expanded out as
\begin{equation} \label{T-vertex}
\Gamma_{\mu}^{T}(p,k,q) = \sum_{i=1}^{8}
\tau_{i}(p^{2},k^{2},q^{2}) T^i_{\mu}(p,k) \,,
 \end{equation}
where the transverse scalar form factors, i.e. the functions
$\tau_i$, remain unknown, and the 8 transverse tensors are
conveniently defined as
 %%%
 \bea \label{T_i} T^1_{\mu}(p,k) &=& p_{\mu}(k \cdot
q)-k_{\mu}(p \cdot q) \,,
\nn \\
T^2_{\mu}(p,k)&=& \left[p_{\mu}(k \cdot q)-k_{\mu}(p \cdot
q)\right](\pslash+\kslash) \,,
\nn \\
T^3_{\mu}(p,k)&=&q^2\gamma_{\mu}-q_{\mu}\qslash \,,
\nn \\
T^4_{\mu}(p,k)&=& q^{2} \left[ \gamma^{\mu} \left( \not\!{k} +
\not\!{p} \right) - \left( k+p \right)^{\mu} \right] \nonumber \\
&& + 2 (k-p)^{\mu} \sigma_{\nu\lambda}\, p^{\nu}k^{\lambda} \,,
\nonumber \\
T^5_{\mu}(p,k)&=&- \sigma_{\mu \nu}\, q^\nu \,,
\nn \\
T^6_{\mu}(p,k)&=&\gamma_{\mu} (p^2-k^2)+(p+k)_{\mu}\qslash \,,
\nn \\
\nn T^7_{\mu}(p,k)&=&\half
(p^2-k^2)\left[\gamma_{\mu}(\pslash+\kslash)-(p+k)_{\mu}\right]
\\ \nn && -(p+k)_{\mu}\sigma_{\nu\lambda}\, p^{\nu}k^{\lambda} \,,
\\
T^8_{\mu}(p,k)&=& \gamma_{\mu}\sigma_{\nu\lambda}\,
p^{\nu}k^{\lambda} -p_{\mu}\kslash+k_{\mu}\pslash \,.
 \eea
Note that these definitions explicitly obey the relations:
 \bea
 q^{\mu} T^i_{\mu}(p,k)=0 \, \quad T_{\mu}(p,p)= 0 \,.
 \eea
It is worth noting that the above tensor basis guarantees a
transverse vertex free of kinematical singularities when $k^2
\rightarrow p^2$. It is slightly different from the initial one
put forward by Ball and Chiu,~\cite{Ball:1980ay}. They carried out
one-loop calculation of the electron-photon vertex in QED in the
Feynman gauge. Guided by this calculation, they proposed the
transverse basis, which ensured the corresponding form factors
were independent of kinematic singularities. However, a later
evaluation of the same vertex in an arbitrary covariant gauge by
K$\imath$zilers\"{u} \textit{et al}.,~\cite{Kizilersu:1995iz}
revealed that a modification of the basis was required to retain
the absence of kinematic singularities for this general case. This
consideration was later also extended to the case of finite
temperature in \cite{Ayala:2001mb}.

In the next sections, we present one-loop perturbative results for
the longitudinal and transverse vertex form factors for some
special kinematics of interest. We employ the singularity free
basis proposed in~\cite{Kizilersu:1995iz}.

%%%%%%%%%%%%%%%%%%%%%%%%%%%%%%%%%%%%%%%%%%%%%%%%%%%%%%%%%%%%%%%%%
%%%%%%%%%%%%%%%%%%%%%%%%%%%%%%%%%%%%%%%%%%%%%%%%%%%%%%%%%%%%%%%%%
\section{The Symmetric Limit}
\label{sec:SymLimit}
%%%%%%%%%%%%%%%%%%%%%%%%%%%%%%%%%%%%%%%%%%%%%%%%%%%%%%%%%%%%%%%%%
%%%%%%%%%%%%%%%%%%%%%%%%%%%%%%%%%%%%%%%%%%%%%%%%%%%%%%%%%%%%%%%%%

Davydychev {\em et. al.}~\cite{Davydychev:2000rt} has provided
one-loop quark-gluon vertex for arbitrary and distinct off-shell
momenta in any gauge and dimensions. It provides us with an
excellent platform to deduce results in different kinematical
limits to provide a practical guide towards its possible
non-perturbative constructions. We can also infer singularity
structure of this vertex and its connection to the multiplicative
renormalizability of the massless quark propagator.

\begin{figure}[h]
\includegraphics[angle=0,width=0.45\textwidth]{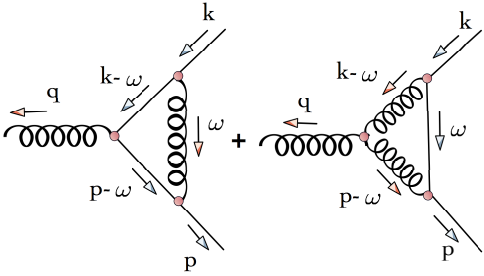}
\vspace{-3.5cm} \caption{The Abelian (left diagram) and
non-Abelian (right diagram) contributions to the one-loop
quark-gluon vertex.} \label{fig1}
\end{figure}

In this section, we present analytical expressions as well as
numerical computations for the longitudinal vertex in the
symmetric case: $p^2=k^2=q^2$, see Fig.~(\ref{fig1}). These
results provide a guide, for this kinematical configuration, to
which all corresponding non-perturbative results should reduce
when the coupling strength is sufficiently weak.

%The general features of the  longitudinal form factors appear
%consistent  in the infra-red. However $\lambda_{2}$ was shown to
%need further study, it tends a constant at zero momentum in
%analytical treatments while it diverges in lattice data \cite{BT}.
%The lattice data for $\lambda_2$ has large errors it suggested
%infrared strength that is seriously underestimate by the model. In
%\cite{MARCE}  the quark gluon vertex was studied using the
%Curci-Ferrari model in this work $\lambda_2$ is a constant. For
%increasing momenta, the most dominant form factor $\lambda_{1}$
%flattens out in lattice studies more than it does in the SDES
%studies.
%In the Landau gauge, longitudinal components of the vertex can
%only be studied at zero gluon momentum, so MOM is the only
%feasible renormalisation scheme.
%Previously \cite{SKW}, the Quark-Gluon vertex in a momentum
%subtraction scheme was studied in the ''symmetric'' momentum
%configuration where $q=-2p$. In this case, all the form factors
%are zero apart from $\lambda_{1}$, $\tau_{3}$ and $\tau_{5}$.

At one-loop perturbation theory, there are two diagrams which
contribute to the longitudinal and transverse components of the
vertex: Abelian ($a$) and non-Abelian ($b$), corresponding to the
left and right diagrams in Fig.~(\ref{fig1}), respectively.

\vspace{-3.0cm}

 \subsection{The Longitudinal Form Factors}

 The longitudinal form factors are defined through
 Eqs.~(\ref{L-Vertex},\ref{L_i}).
 $\lambda_{1}(p^{2},k^{2},q^{2},1/\epsilon)$ is the only one of these which
 is UV-divergent at one-loop. Note that the space-time dimension
 is defined as $n=4-2 \epsilon$; $n \rightarrow 4$ as $\epsilon \rightarrow
 0$. We employ momentum subtraction (MOM) renormalization scheme to
 define the renormalized vertex (identified by the subscript $R$ below), such that at a large enough momentum
 scale $p^2=-\mu^2$, tree-level perturbation theory is valid and hence, for the
 symmetric case,
 \bea
   \Gamma^{\mu}_R(p^{2},p^{2},p^{2})|_{p^{2}= -\mu^{2}} \equiv \Gamma^{\mu}_R(p^{2},-\mu^2)|_{p^{2}= -\mu^{2}}
   = \gamma^{\mu} \,.
 \eea
 This renormalization condition translates as:
 \begin{eqnarray}
  \lambda_{1R}(p^{2},p^{2},p^{2})_{p^{2}=-\mu^{2}} \equiv \lambda_{1R}(p^{2},-\mu^2)|_{p^{2}=-\mu^{2}} =
  0 \,,
 \end{eqnarray}
 and determines the vertex renormalization constant $Z_{1F}^{-1}(\mu^{2},\varepsilon)$ as
 follows:
 \begin{eqnarray}
 \Gamma^{\mu}_R(p^{2},-\mu^2)=Z_{1F}^{-1}(\mu^{2},\varepsilon)\Gamma_B^{\mu}(p^2,\varepsilon)
 \,,
 \label{renormalization}
 \end{eqnarray}
 where the subscript $B$ specifies the bare quantities.
 In one-loop perturbation theory,
\begin{eqnarray}
Z_{1F}(\mu^{2},\varepsilon)=1+\lambda^B_{1}(-\mu^{2},\varepsilon)
\,, \label{renormalization cte}
 \end{eqnarray}
where the bare quantities depend upon the momentum scale $p^2$ and
on the regulator $\epsilon$, having the pole divergence
$1/\epsilon$ for $\epsilon \rightarrow 0$. We convert the bare
coupling into the renormalized one through the prescription: $
{g^2}/{4 \pi} = \alpha(\mu) \, {\cal Z_{\alpha} }(\mu^2,
  \epsilon).$
Note that as $
  {\cal Z_{\alpha} }(\mu^2, \epsilon) = 1 + {\cal O}(\alpha)$,
 we can write $
  {g^2}/{4 \pi} = \alpha(\mu)$ to the one-loop order.
Therefore, explicitly
 \bea
 Z_{1F}= 1 + \frac{1}{\epsilon}
 \Big[ (1-\xi) C_a + \frac{3}{4} (2 -\xi) C_b
 \Big] + {\rm Fin}_{Z}  \,, \\ \nn
 \eea
where
\begin{eqnarray}
 C_{a}&=& \frac{\alpha(\mu)}{4\pi} \, \left(C_{F}-\frac{1}{2}C_{A}
 \right) \,, \nn \\
  C_{b} &=& \frac{\alpha(\mu)}{4\pi} \, C_{A}
 \,. \nn
\end{eqnarray}
Note that $C_F=(N^2-1)/2N$ is the eigenvalue of the Casimir
operator in the fundamental representation of $SU(N)$, while
$C_A=N$ is that in its adjoint representation. The term
proportional to $C_a$ corresponds to the Abelian QED-like diagram
and the term involving $C_b$ to the non-Abelian triple-gluon
contribution,~\cite{Celmaster:1979km}. ${\rm Fin}_{Z}$ is the
finite part of the renormalization constant $Z_{1F}$ and is given
by:
\begin{eqnarray}
{\rm Fin}_{Z}&=& C_{a} \left( 1-\xi \right) \bigg\{ \frac{ m^4 -
\mu^4 }{\mu^4} L(-\mu^2) - \frac{m^2}{\mu^2} + C_m  \bigg\} \nonumber \\
&&  \hspace{-1cm} - \frac{1}{4} C_{b} (2 -\xi) \bigg\{ - \frac{
m^4 - m^2 \mu^2 - 2 \mu^4 }{\mu^4} L(-\mu^2) + \frac{m^2}{\mu^2}
\nn \\
&&  \hspace{-1cm} - \ln \left( \frac{m^2}{\mu^2} \right) - (m^2 +
\mu^2) \varphi_{1}(-\mu^2) -2 - 3 C_m \bigg\} \,. \label{FinZ} \,
\end{eqnarray}
where $C_m = 1 - \ln(m^2) + \ln(4\pi) - \gamma_{E}$, $\gamma_{E}$
being the Euler constant, and $L(p^2)= \ln(1- {{p}^{2}}/{m^{2}})$.
Moreover, we have employed the following convention for the
three-point integrals:
\begin{eqnarray}
 && \hspace{-.5cm} i \pi^{n/2} \varphi_1(k^2,p^2,q^2) = \hspace{-.2cm} \int \frac{d^nw}{(w^2 - m^2) (k-w)^2 (p-w)^2 },
 \nn \\ \nn \\
 && \hspace{-.5cm} i \pi^{n/2} \varphi_2(k^2,p^2,q^2) = \hspace{-.2cm} \int \frac{d^nw}{w^2 [(k-w)^2-m^2] [(p-w)^2-m^2] },
 \nn
\end{eqnarray}
and $\varphi_{1,2}(p^2) \equiv \varphi_{1,2}(p^2,p^2,p^2)$.
Explicitly,
\begin{eqnarray}
&& \varphi_1(p^2) = \frac{1}{p^2 \sqrt{3}} \bigg\{ 2 {\rm
Cl}_{2}\left( \frac{\pi}{3} \right) + 2 {\rm Cl}_{2} \left(
\frac{\pi}{3} + 2 \theta \right)  \nn \\
&& \hspace{2.6cm} + {\rm Cl}_{2} \left( \frac{\pi}{3} - 2 \theta
\right) + {\rm Cl}_{2} (\pi - 2 \theta) \bigg\}, \nn \\
&&  \varphi_2(p^2) = \frac{2}{p^2 \sqrt{3}} \bigg\{ 2 {\rm
Cl}_{2}\left( \frac{2 \pi}{3} \right) + {\rm Cl}_{2} \left(
\frac{\pi}{3} + 2 \tilde{\theta} \right) \nn \\
&& \hspace{2.6cm} +  {\rm Cl}_{2} \left( \frac{\pi}{3} - 2
\tilde{\theta} \right) \bigg\} \,, \nn
\end{eqnarray}
where ${\rm Cl}_{2}$ are the Clausen functions with $\theta$ and
$\tilde{\theta}$ defined as (see C(20-22) of
\cite{Davydychev:2000rt})
\begin{eqnarray}
\tan \theta = \frac{ p^2 - 2 m^2 }{ p^2 \sqrt{3} } \,, \quad \tan
\tilde{\theta} = \sqrt{\frac{ p^2 - 4 m^2 }{ 3 p^2 }} \,.\nn
\end{eqnarray}
\noindent In terms of these definitions, our evaluated analytical
results for the longitudinal form factors in the symmetric limit
are enlisted as follows.

\begin{widetext}
\noindent{\bf Abelian contribution for $\lambda$'s:}
\begin{eqnarray}
\lambda_{1R}^{a}(p^{2},-\mu^2)
&=&\frac{C_{a}(\xi-1)}{p^{4}\mu^{4}}\bigg\{p^{4}(m^{4}-\mu^{4})L(-\mu^2)+\mu^{4}(p^{4}-m^{4})L(p^2)
-m^{2}p^{2}\mu^{2}(p^{2}+\mu^{2} )\bigg\} \,,
\label{lamda1a} \\
 \lambda_{2}^{a}(p^{2})&=& \frac{C_{a}(\xi-1)}{2p^6}
\bigg\{p^2(2m^{2}+p^{2})+2m^4L(p^2) \bigg\} \,,
\label{lambda2a} \\
\lambda_{3}^{a}(p^{2})&=&\frac{C_{a}(\xi-4)m}{p^{4}}
 \bigg\{ p^{2}+m^{2}L(p^2) \bigg\} \,. \label{lambda3a}
\end{eqnarray}
\noindent{\bf Non-Abelian contribution for $\lambda$'s:}
\begin{eqnarray}
 \lambda_{1R}^{b}(p^{2},-\mu^2) &=& \frac{C_{b}(\xi-2)}{4 p^4
\mu^4} \bigg\{  p^4 (m^{4} - m^{2} \mu^{2} - 2\mu^{4}) L(-\mu^{2})
- \mu^4 (m^{4}+ m^{2} p^{2} - 2 p^{4})
L(p^{2}) + p^4 \mu^4  \ln{ \left(- \frac{p^{2}}{\mu^{2}} \right)} \nonumber \\
&& \hspace{1.7cm}  - \mu^2 p^2 \left[ (\mu^2 + p^2) m^2 - \mu^2
p^2 (m^2 + \mu^2) \varphi_1 (-\mu^2) + \mu^2 p^2(m^2 - p^2)
\varphi_1 (p^2) \right] \bigg\} \,,  \label{lambda1b} \\
\lambda_{2}^{b}(p^{2})&=&\frac{C_{b}}{24 p^6}
\bigg\{(2+\xi)p^{2}\left[ p^2 (2 m^2 - p^2) \varphi_1(p^2) + 3 p^2
- 2 p^2 \ln\left(- \frac{p^{2}}{m^{2}} \right) + (m^2 + 2 p^2)
L(p^2)
 \right]   \nn \\
&&\hspace{2.5cm} - 3(2-\xi)
\left[p^{2}(p^{2}+2m^{2})+2m^{4}L(p^2)\right]\bigg\} \,,
\label{lambda2b}  \\
\nonumber
\lambda_{3}^{b}(p^{2})&=&\frac{C_{b}m}{8p^4[p^{4}+m^{4}-m^{2}p^{2}]}
\bigg\{\bigg[2(\xi-6)m^{6}+3(\xi-4)m^{2}p^{2}(p^{2}-m^{2})+\xi
p^{6} \bigg]L(p^2)
\\&& \hspace{3.7cm} + p^{2}\bigg[2(\xi-6)(p^{4}+m^{4}-m^{2}p^{2})- \xi
p^{4}\ln\left(-\frac{p^{2}}{m^{2}}\right)\bigg] \bigg\} \,.
\label{lambda3b}
\end{eqnarray}

\end{widetext}

\begin{figure}[h!]
\includegraphics[angle=0,width=0.5\textwidth]{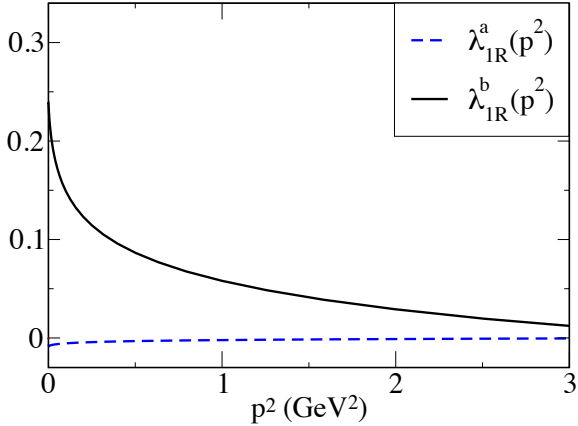}
\vspace{-0.7cm} \caption{One-loop form factors
$\lambda_{1R}^{a,b}$ in the Feynman gauge.} \label{fig2}
\end{figure}

 Charge conjugation symmetry of the quark-gluon
 vertex implies:
 $\lambda_{4}(p^2,k^2,q^2)=-\lambda_{4}(k^2,p^2,q^2)$. Therefore,
 it is naturally zero in the symmetric limit $k^2=p^2$.

 Although the perturbative result is valid only for large
$p^2$, we have taken the liberty to extrapolate it into the
infrared for a comparative analysis and in the hunt for possible
singularity structure. We can analytically calculate
$\lambda_i^{a,b}(p^2=0)$. In the Landau gauge,
 \bea
 \lambda_{1 \, R}^{a} (p^{2}=0) &=& \lambda_{2}^{a} (p^{2}=0) = \lambda_{4}^{a,b} (p^{2}=0) = 0 \,, \nonumber \\
 \lambda_{3}^{a} (p^{2}=0) &=& \frac{3 C_a}{ 2 m } \rightarrow -0.020/{\rm
 GeV}
   \,, \nonumber  \\
 \lambda_{1 \, R}^{b} (p^{2}=0) &=& -\frac{C_b}{ 8 \mu^4 } \Big\{
\mu^2 ( 5 \mu^2-2m^2 ) \nonumber \\
&& \hspace{1cm} +2 \mu^2 (m^2 +\mu^2) \varphi_{1}(-\mu^2)
\nonumber \\
&& \hspace{1cm}  + 2 (m^4 -m^2 \mu^2 -2 \mu^4)   L(-\mu^2) \nonumber \\
&& \hspace{1cm} -2 \mu^4 \ln\left( \frac{\mu^2}{m^2} \right)
\Big\} \rightarrow
 .119 \,, \nonumber \\
 \lambda_{2}^{b} (p^{2}=0) &=& \frac{C_b}{ 12 m^2 }\rightarrow
 .177/{\rm GeV}^2
  \,, \nonumber \\
 \lambda_{3}^{b} (p^{2}=0) &=& \frac{3 C_b}{ 4 m } \rightarrow .183/{\rm
 GeV}
   \,, \nn
\end{eqnarray}
where the numerical values have been evaluated in QCD for
$\alpha=.118$, $m=.115$ GeV and $\mu=2$ GeV.

\begin{figure}[h!]
\includegraphics[angle=0,width=0.5\textwidth]{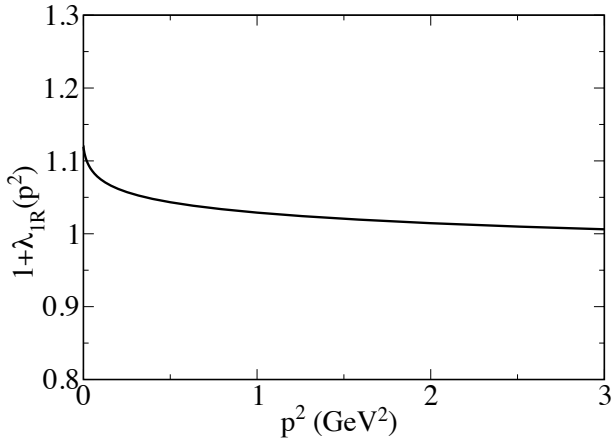}
\vspace{-0.5cm} \caption{Full one-loop form factor
$1+\lambda_{1R}$ in the Landau gauge.} \label{fig3}
\end{figure}

There are several observations in place:

\begin{enumerate}

\item

$\lambda_{1R}^{a}(p^{2},-\mu^2)$ and $\lambda_{2}^{a}(p^{2})$
identically vanish in the Landau gauge. This is consistent with
the fact that so does the wavefunction renormalization $F(p^2)$ in
the same gauge.

\begin{figure}[h!]
\includegraphics[angle=0,width=0.5\textwidth]{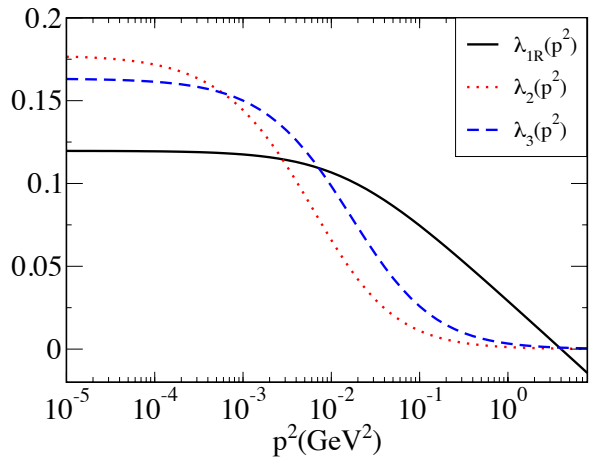}
\vspace{-0.5cm} \caption{One-loop form factors $\lambda_{1R}$,
$\lambda_{2}$, and $\lambda_{3}$ in the Landau gauge. We draw them
till deep infrared to show that all of them saturate in that
limit.} \label{fig4}
\end{figure}

\begin{figure}[h!]
\includegraphics[angle=0,width=0.5\textwidth]{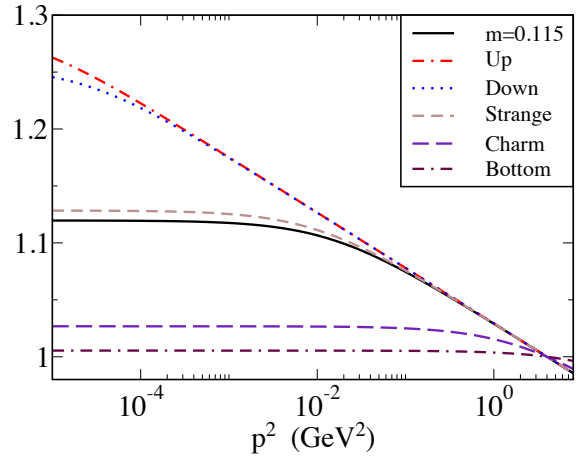}
\vspace{-0.5cm} \caption{One-loop form factor $1+\lambda_{1R}$ in
the Landau gauge for different quark masses. Smaller masses imply
larger infrared enhancement just as in full non-perturbative QCD.
All masses are given in GeV.} \label{fig5}
\end{figure}

\item

Therefore, for the sake of comparison, we plot
$\lambda_{1R}^{a}(p^{2},-\mu^2)$ and
$\lambda_{1R}^{b}(p^{2},-\mu^2)$ in the Feynman gauge, where both
are explicitly non-zero. Note that the non-Abelian contribution is
positive in the infrared and its magnitude there is about 95\%
more enhanced, see Fig.~(\ref{fig2}).

\begin{figure}[h!]
\includegraphics[angle=0,width=0.5\textwidth]{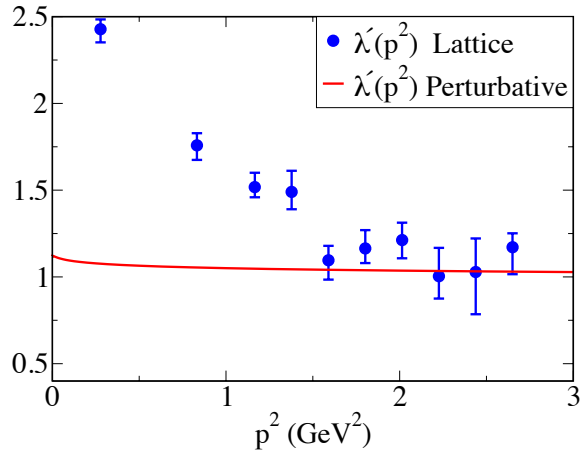}
\vspace{-0.5cm} \caption{One-loop form factor
$\lambda'(p^2)=\lambda_{1R}(p^2) + p^2 \tau_3(p^2)$ in the
symmetric limit in the Landau gauge and its comparison with
lattice results,~\cite{Skullerud:2004gp}.} \label{fig6}
\end{figure}

\item

 Note that even the one-loop calculation shows an infrared
enhancement of $1+\lambda_{1R}(p^{2},-\mu^2)$, see
Fig.~(\ref{fig3}), plotted in the Landau gauge, $\xi=1$. However,
expectedly this is only a small increase as compared to the
non-perturbative effect observed in lattice and SDE studies, see
for example~\cite{Skullerud:2003qu,Bhagwat:2004kj}. One-loop
result is responsible for about 10\% infrared increase of
$1+\lambda_{1R}(p^{2},-\mu^2)$ from its tree-level value, while
the non-perturbative effects reveal more than a 100\% rise. These
numbers are at the lowest momentum value where lattice has
computed its results,~\cite{Skullerud:2003qu}.

\begin{figure}[h!]
\includegraphics[angle=0,width=0.5\textwidth]{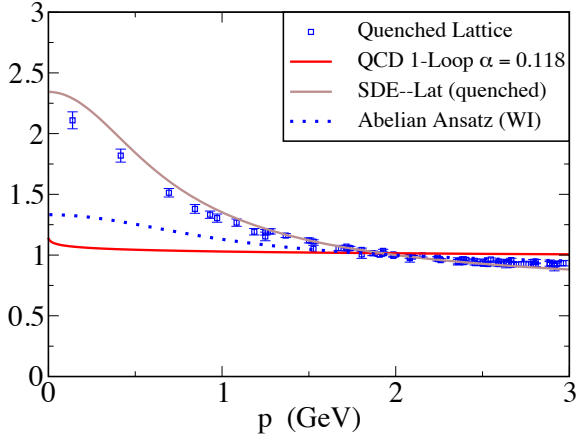}
\vspace{-0.5cm} \caption{One-loop form factor
$1+\lambda_{1R}(p^2,p^2,0)$ in the soft photon limit in the Landau
gauge and its comparison with lattice as well as SDE results.}
\label{fig7}
\end{figure}

\item

We also check for the deep infrared behavior of one-loop
$\lambda_{1R}(p^{2},-\mu^2)$. A simple analytical and numerical
check shows that this form factor saturates in this limit at the
value $\sim 0.119$. The same qualitative feature is also observed
for $\lambda_{2}(p^{2})$ and $\lambda_{3}(p^{2})$, all plotted in
Fig.~(\ref{fig4}). There is no infrared divergence in the
symmetric case. This should be considered as a guideline for
lattice studies for which we expect $p^2 \lambda_{2}(p^{2})$ to
vanish in the infrared. Any non-zero value will imply a divergent
infrared $\lambda_{2}(p^{2})$,~\cite{Skullerud:2003qu}.

\item

Variation of current mass for the quarks shows that the major
infrared enhancement is for the lightest quarks and it diminishes
with the increase of current quark mass. Quantitatively, the rate
of decrease goes as $\sim 7.2\% \rightarrow 7 \% \rightarrow 2.2
\% \rightarrow 0.3\%$ as we go from $u,d \rightarrow s \rightarrow
c \rightarrow b$, see Fig.~(\ref{fig5}).

\item

  In Figs.~(\ref{fig6},\ref{fig7}), we compare one-loop results
against the lattice as well as
SDE-results,~\cite{Skullerud:2004gp,Skullerud:2003qu,Bhagwat:2004kj}.
As mentioned earlier, the perturbative rise is only a very small
percentage of the non-perturbative effects. We use $m=0.115$ GeV
and $m=0.06$ GeV, respectively in Fig.~(\ref{fig6}) and
Fig.~(\ref{fig7}), to make direct comparison with the lattice
results.

\end{enumerate}

\subsection{The Transverse Form Factors}

On the other hand, the transverse vertex is defined via
Eqs.~(\ref{T-vertex},\ref{T_i}).  Just as for
$\lambda_4(k^2,p^2,q^2)$, the symmetry of the quark-gluon vertex
under the interchange of quark and anti-quark requires
 \bea
  \tau_4(k^2,p^2,q^2) &=& - \tau_4(p^2,k^2,q^2) \,, \nn \\
  \tau_6(k^2,p^2,q^2) &=& - \tau_6(p^2,k^2,q^2) \,.
 \eea
Therefore, in the symmetric limit, both of these form factors
vanish identically. We now present explicit analytical results for
the symmetric limit, and then carry out a numerical analysis.
Moreover, we adopt the simplified notation
$\tau_i^{a,b}(p^2,p^2,p^2) \equiv \tau_i^{a,b}$, and write out the
$\tau_i^{a,b}$ explicitly as follows:

\begin{widetext}
\noindent{\bf Abelian contribution for $\tau$'s:}
\begin{eqnarray}
\nn \tau_1^a &=& \frac{2C_a  m}{3p^6} (\xi-4)\left\lbrace p^2[3-(2m^2+p^2)
\varphi_2(p^2) -4f(p^2)]+(m^2+2p^2)L(p^2)\right\rbrace \,,\\
\nn \tau_2^ a&=&\frac{ C_a} {3 p^6}\bigg\{- 4 m^2(\xi -1)-p^2(\xi -5 )+
\left[4 m^4 (\xi -1)-2 p^4\right]\varphi_2(p^2)+4(\xi -1)\left(2 m^2
+p^2\right) f(p^2)\\
\nn &&+2  \left[2 m^2-(\xi -1) p^2\right]L\left(p^2\right)\bigg\} \,, \\
\nn \tau_3^a &=& -  \frac{C_a}{3p^{6}} \bigg\{ 3 p^{2} \left[ (\xi
+1) p^{2} -(\xi -1) m^{2} \right]+ p^{2} \left[ 2 (\xi-1) m^{4} +
4 m^{2} p^{2} + (1 -2 \xi) p^{4} \right]
  \varphi_{2}(p^{2})  \\
\nn && - 2 p^{2} \left[ (\xi +1) p^{2} -2 (\xi -1) m^{2} \right]
f(p^2) - (m^2-p^{2})
\left[ (\xi +1) p^{2} +(\xi -1) m^{2} \right] L(p^{2}) \bigg\} \,, \\
\nn \tau_4^a &=&  0 \,, \\ \nn \\
\nn \tau_5^a &=& -\frac{C_a  m \xi}{3p^4} \left\lbrace
p^2[3+2(m^2-p^2) \varphi_2(p^2)
+4f(p^2)]+(5m^2-2p^2)L(p^2)\right\rbrace \,,\\ \nn \\
\nn \tau_6^a &=&0 \,, \\ \nn \\
\nn \tau_7^a &=&\frac{2 C_a m
\xi}{3 p^6} \left\lbrace p^2 [1 + (2 m^2-p^2)   \varphi_2(p^2) + 4
f(p^2)] + (3 m^2 - 2 p^2)
L(p^2)\right\rbrace \,, \\
%\nn \tau_7^a &=&\frac{2 C_a m \xi }{3 p^6} \bigg\{4 p^2 f(p^2)+p^2
%\left(2 m^2 \varphi_2-p^2
%\varphi_2+1\right)+\left(3 m^2-2 p^2\right)L(p^2)\bigg\} \,, \\
\tau_8^ a&=&-\frac{2 C_a}{3 p^4} \bigg\{ p^2 \left(2
m^2+p^2\right)\varphi_2(p^2)+4p^2f(p^2)+2(m^2-p^2) L(p^2)\bigg\}
\,. \label{TausAbelianSymLimit}
\end{eqnarray}

\noindent{\bf Non-Abelian contribution for $\tau$'s:}
\begin{eqnarray}
\nn \tau_1^b &=&\frac{C_b \ m}{24p^6(m^4-m^2p^2+p^4)^2} \bigg\lbrace p^2\Big[8m^8(\xi-6)+4m^6p^2(24-3\xi-\xi^2)\\
\nn && -6m^4p^4(24-3\xi -2\xi^2) +2m^2p^6(48-5\xi -3\xi^2)-2p^8(24-3\xi -4\xi^2)\Big] \ln\left( -\frac{p^2}{m^2} \right) \\
\nn && +2p^2 (m^4-m^2p^2+p^4)\Big[12m^2(m^2-p^2)(\xi-3)-3p^4(12-4\xi+\xi^2)\\
\nn && +2(m^4-m^2p^2+p^4)[2m^2(\xi-6)+p^2(6-2\xi+\xi^2)]\varphi_1(p^2)\Big]\\
\nn && +2\Big[4m^{10}(2\xi-3)-m^8p^2\xi(13+2\xi)+2m^6p^4(6+7\xi+4\xi^2)-m^4p^6(48+\xi+6\xi^2)\\
\nn && +2m^2p^8(18-2\xi-\xi^2)-p^{10}(24-5\xi-2\xi^2)\Big] L(p^2)
\bigg\rbrace \,, \nn \\
\nn \tau_2^ b&=&-\frac{C_b}{12 p^6 \left(m^4-m^2 p^2+p^4\right)}\bigg\{ 4 p^2\left(m^4-m^2 p^2+p^4\right) -2 (\xi -2)(\xi p^2-2 m^2) \left(m^4-m^2 p^2+p^4\right)   \\
\nn && -\Big[4 (\xi -2) m^6+8 m^4 p^2+\left(\xi ^2-8\right) m^2 p^4+4 \xi p^6\Big] \ln \left(-\frac{p^2}{m^2}\right)\\
\nn && + \Big[4(\xi -2) m^4+2 (\xi +2) m^2 p^2+(\xi -2) (\xi +1) p^4\Big]\left(m^4-m^2 p^2+p^4\right) \varphi_1(p^2)\\
\nn && + \Big[(-2 \xi^2+5\xi-6 ) m^6+m^4 p^2(2 \xi^2-\xi +6) +(\xi
-6) m^2 p^4+4 \xi p^6\Big]L\left(p^2\right)\bigg\} \,, \\
\nn \tau_{3}^{b} &=& \frac{C_b}{ 24p^{6} ( m^{4} - m^{2} p ^{2} + p^{4} ) } \bigg\{ 6 p^{2} (m^{4} - m^{2} p ^{2} + p^{4}) \Big[ (2 + 3 \xi - \xi^{2}) p^{2} + (\xi-2) m^{2} \Big] \\
\nn && - p^{2} \Big[ 4 (\xi -2) m ^{6} - 2 (\xi^{2} - 3 \xi - 6)
m^{4} p^{2} + (5 \xi^{2} - 18 \xi - 12) m^{2} p^{4} - 2 (\xi^{2} -
5 \xi - 2) p^{6} \Big] \ln \left( - \frac{p^{2}}{m^{2}} \right) \\
\nn &&+ p^{2} (m^{4} - m^{2} p ^{2} + p^{4}) \Big[ \xi^{2} p^{2}
(p^{2} -2 m^{2}) + 4 \xi (m^{2} + p^{2})^{2}
- 8 (m^{4} - m^{2} p^{2} + p^{4}) \Big] \varphi_{1}(p^{2})\\
\nn && + 2 (m^{2} - p^{2}) \Big[ (\xi-2) m^{6} - 2 \xi (\xi -3)
m^{4} p^{2} - \xi (\xi - 6) m^{2} p^{4} + (\xi^{2} - 5 \xi - 2)
p^{6}
\Big] L(p^{2}) \bigg\} \,, \\
\nn \tau_4^b &=& \tau_6^b =0 \,, \\
\nn \tau_5^b &=&-\frac{C_b \ m}{12p^4(m^4-m^2p^2+p^4)}\bigg\lbrace  p^2\xi\Big[2m^4-m^2p^2(\xi-4)+2p^4(\xi-2)\Big]\ln\left( -\frac{p^2}{m^2} \right)\\
\nn && + p^2(m^4-m^2p^2+p^4)\Big[6\xi+[p^2(18-8\xi+\xi^2)-2m^2\xi]\varphi_1(p^2)\Big]\\
\nn && +2\xi(m^2-p^2)\Big[4m^4+m^2p^2(\xi-4)+p^4(\xi-2)\Big]L(p^2)\bigg\rbrace \,, \\
\nn \tau_7^b &=&\frac{C_b m \xi}{12 p^6 (m^4 - m^2 p^2 +
     p^4)^2 } \bigg\lbrace -p^2 (m^4 - m^2 p^2 + p^4) [-2 m^4 + 2(1-\xi) m^2 p^2 + p^4 (\xi-2)]\\
   \nn && -
     p^2 [4 m^8 - 6 m^6 p^2 + 8 m^4 p^4-
        m^2 p^6 (4 + \xi) + 2 p^8 (1 + \xi) ] \ln\left(-\frac{p^2}{m^2}\right)\\
   \nn && - p^2 (m^4 - m^2 p^2 + p^4) [4 m^6 - 4 m^4 p^2 +
   4 m^2 p^4] \varphi_1(p^2) \\
        \nn && +
     2 (m^2- p^2) [3 m^8  +
        m^6 p^2 (\xi -6)- m^4 p^4 (\xi -7) + 2 m^2 p^6 (\xi-2) +
        p^8 (\xi +1)] L(p^2)\bigg\rbrace \,, \\
\nn \tau_8^b &=& \frac{C_b}{12p^{4} (m^{4} - m^{2} p^{2}+ p^{4})}
\bigg\{ p^{2} \Big[ 2 m^4(\xi - 6) - (\xi^{2} - 2 \xi - 12) m^{2}
p^{2} +2p^4 (\xi^{2} - 3\xi - 6)\Big] \ln\left( -
\frac{p^{2}}{m^{2}} \right) \\
\nn &&+ p^{2} (m^{4} - m^{2} p ^{2} + p^{4}) \Big[ (\xi^{2} - 6
\xi + 12) p ^{2} - 2 (\xi -6) m ^{2} \Big]
\varphi_{1}(p^{2})  \\
&& + 2 (m^{2} - p^{2}) \Big[ (\xi - 6) m^{4} + (\xi^{2} - 5 \xi +
6) m^{2} p^{2} + (\xi^{2} - 3\xi - 6) p^{4} \Big] L(p^{2}) \bigg\}
\,, \label{TausNonAbelianSymLimit}
\end{eqnarray}
\end{widetext}
where function $f(p^2)$ is defined as
\begin{eqnarray}
f(p^2) = \left\{
\begin{array}{ll} \sqrt{\frac{p^2-4m^2}{4p^2}}
\ln\frac{\sqrt{p^2-4m^2}+\sqrt{p^2}}{\sqrt{p^2-4m^2}-\sqrt{p^2}}
\; , &
\;\;\; p^2>4m^2 \; , \\ \\
\sqrt{\frac{4m^2-p^2}{p^2}} \arctan\sqrt{\frac{p^2}{4m^2-p^2}} \;
, & \;\;\; p^2<4m^2 \; .
\end{array} \right.
\end{eqnarray}

%\begin{eqnarray}
%&& \hspace{-0.6cm} \label{f} f(p^2) =  \sqrt{\frac{p^2-4m^2}{4
%p^2}} \ln\frac{\sqrt{p^2-4m^2} + \sqrt{p^2}}{\sqrt{p^2-4m^2} -
%\sqrt{p^2}}
%\; \; \; \; p^2>4m^2 \,, \nn \\
%&& \hspace{0.3cm} \label{f}   = \sqrt{\frac{4m^2-p^2}{p^2}}
%\arctan \sqrt{\frac{4m^2-p^2}{p^2}} \qquad \; p^2<4m^2 \,. \nn
%\end{eqnarray}

\begin{figure}[h!]
\centering
\includegraphics[width=0.5\textwidth]{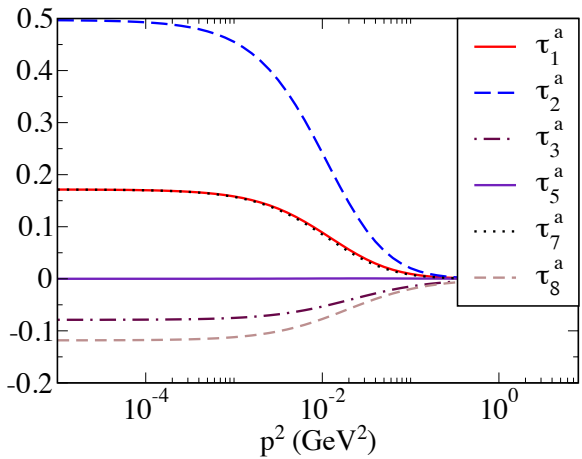}
\caption{Abelian $\tau_i^a(p^2)$ for the symmetric configuration
of momenta $k^2=p^2=q^2$.} \label{fig8}
\end{figure}

In the Landau gauge, numerical results for the Abelian components
of the symmetric transverse form factors,
Eqs.~(\ref{TausAbelianSymLimit}), are presented in
Fig.~(\ref{fig8}). One of the checks of their numerical evaluation
is the deep infrared behavior: $\tau_i^a(p^2=0)$, which can be
calculated analytically:
\begin{eqnarray}
\tau_{1}^{a} (p^{2}=0) &=& - \frac{C_a}{ 6 m^{3} }  \rightarrow .171 / {\rm GeV}^{3}  \,, \nonumber \\
\tau_{2}^{a} (p^{2}=0) &=& - \frac{C_a}{ 18 m^{4} }  \rightarrow .497 / {\rm GeV}^{4} \,, \nonumber \\
\tau_{3}^{a} (p^{2}=0) &=&   4 \frac{C_a}{ 6 m^{2} }  \rightarrow -.078 / {\rm GeV}^{2} \,, \nonumber \\
\tau_{4}^{a} (p^{2}=0) &=& \tau_{6}^{a} (p^{2}=0) =0  \,, \nonumber \\
\tau_{5}^{a} (p^{2}=0) &=& 0 \,, \nonumber \\
\tau_{7}^{a} (p^{2}=0) &=& - \frac{C_a}{ 6 m^{3} }  \rightarrow .171 / {\rm GeV}^{3} \,, \nonumber \\
\tau_{8}^{a} (p^{2}=0) &=&  \frac{C_a}{  m^{2} }  \rightarrow
-.118 / {\rm GeV}^{2} \,, \label{TausAbelianSymLimitDeepIR}
\end{eqnarray}
where, as before, the numerical values have been evaluated in QCD
for $\alpha=.118$ and $m=.115$ GeV. These values match with the
numerical computation of the plots displayed in Fig.~(\ref{fig8})
in the infrared limit. Note that all the $\tau_i^a$ converge to
finite values in this limit. Therefore, we expect that for
symmetric configuration of momenta, any QED construction (which is
basically the Abelian version of QCD) of the three-point vertex
should not be singular in the infrared limit.

\begin{figure}[h!]
\centering
\includegraphics[width=0.5\textwidth]{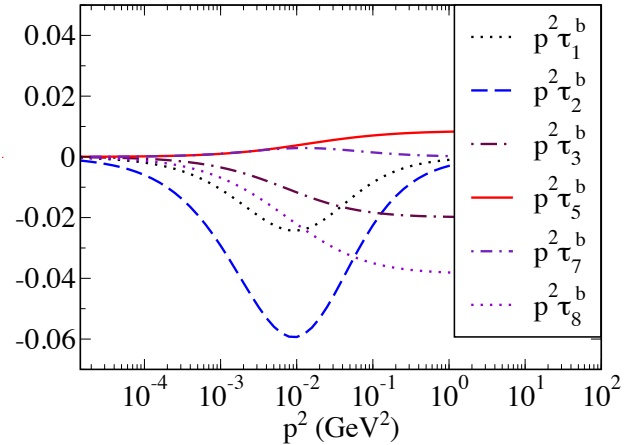}
\caption{Non-Abelian $\tau_i^b(p^2)$, weighted with $p^2$ for the
symmetric configuration of momenta $k^2=p^2=q^2$.} \label{fig9}
\end{figure}

For the non-Abelian transverse coefficients,
Eqs.~(\ref{TausNonAbelianSymLimit}), the deep infrared limit,
$\tau_i^b(p^2 \rightarrow 0)$, in the Landau gauge reads as
\begin{eqnarray}
\tau_{1}^{b} (p^{2} \rightarrow 0) &=& \frac{C_{b}}{18 m^{3}}
 \left[-4 + 3 \ln\left( \frac{p^{2}}{m^{2}} \right) \right] \,,
 \nonumber\\
\tau_{2}^{b} (p^{2} \rightarrow 0) &=& \frac{C_{b}}{24 m^{4}}
 \left[1 + 2 \ln\left( \frac{p^{2}}{m^{2}} \right) \right] \,,
 \nonumber\\
\tau_{3}^{b} (p^{2} \rightarrow 0) &=& \frac{C_{b}}{144 m^{2}}
 \left[-35 + 78 \ln\left( \frac{p^{2}}{m^{2}} \right) \right] \,,
 \nonumber\\
\tau_{4}^{b} (p^{2} \rightarrow 0) &=& 0 \,,
 \nonumber\\
\tau_{5}^{b} (p^{2} \rightarrow 0) &=& -\frac{C_{b}}{8 m}
 \left[-7 + 10 \ln\left( \frac{p^{2}}{m^{2}} \right) \right] \,,
 \nonumber\\
\tau_{6}^{b} (p^{2} \rightarrow 0) &=& 0 \,, \nonumber\\
\tau_{7}^{b} (p^{2} \rightarrow 0) &=& \frac{C_{b}}{12 m^{3}}  \,,
\nonumber\\
\tau_{8}^{b} (p^{2} \rightarrow 0) &=& \frac{C_{b}}{8 m^{2}}
 \left[1 + 10 \ln\left( \frac{p^{2}}{m^{2}} \right) \right] \,.
 \nonumber
\end{eqnarray}
The above results reveal a logarithmic divergence for the
non-Abelian coefficients in the deep infrared regime, $p^2
\rightarrow 0$, which is absent in the Abelian counterpart.
However, this is not in conflict with the requirement of no
kinematical singularities because in the symmetric case, one also
takes the photon momentum squared $ \rightarrow 0$, which is the
dynamical limit of taking the photon to be on-shell. In
Fig.~(\ref{fig9}), we can see that the product $p^2 \tau_i^b
(p^2)$ is well-behaved for infrared momenta. Therefore, this is
what we opt to plot. Note that for several $\tau_i^b$, the factor
$p^2$, necessary to suppress the logarithmic divergence, comes
right from the tensor basis, Eqs.~(\ref{T_i}).

We could think of redefining the transverse tensors to get rid of
the divergence in some of the $\tau_i^b(p^2)$ in the symmetric
limit, but it is not possible to get rid of the overall divergence
arising from the three-gluon vertex configuration. So we resist
the temptation to do it.

With all the results of this section, we have a complete guideline
for any non-perturbative construction of the quark-gluon vertex to
reduce to in the weak coupling regime of the symmetric
configuration of momenta. Moreover, we have analyzed the infrared
singularity structure of each component of the transverse vertex.
We now focus our attention on the asymptotic limit of momenta,
which has played a crucial role in implementing MR of the massless
electron propagator.

%%%%%%%%%%%%%%%%%%%%%%%%%%%%%%%%%%%%%%%%%%%%%%%%%%%%%%%%%%%%%%%%%
%%%%%%%%%%%%%%%%%%%%%%%%%%%%%%%%%%%%%%%%%%%%%%%%%%%%%%%%%%%%%%%%%
\section{The Asymptotic Limit}
\label{sec:AsympLimit}
%%%%%%%%%%%%%%%%%%%%%%%%%%%%%%%%%%%%%%%%%%%%%%%%%%%%%%%%%%%%%%%%%
%%%%%%%%%%%%%%%%%%%%%%%%%%%%%%%%%%%%%%%%%%%%%%%%%%%%%%%%%%%%%%%%%

From the works in QED, we already know that the intricate
structure of the quark-gluon vertex dictates MR of the electron
and hence ensures LKFT for the 2-point function are satisfied.
Brown and Dorey,~\cite{Brown:1989hy}, argue that an arbitrary
construction of the electron-photon vertex does not satisfy the
requirement of MR. It was realized that neither the bare vertex
nor the Ball-Chiu-vertex,~\cite{Ball:1980ay}, which satisfies the
Ward-Fradkin-Green-Takahasi identity
(WFGTI),~\cite{Ward:1950xp,Fradkin:1955jr,Green:1953te,Takahashi:1957xn},
were good enough to fulfill the demands of MR. Since then,
starting from the pioneering work by Curtis and
Pennington,~\cite{Curtis:1990zs}, there have been improved
attempts to incorporate the implications of LKFT in constructing a
reliable electron-photon vertex {\em
Ansatz},~\cite{Dong:1994jr,Bashir:1994az,Bashir:1995qr,Bashir:2007qq,Kizilersu:2009kg,Bashir:2011vg,Bashir:2011dp,Binosi:2016wcx,Pennington:2016vxv}.
This owes itself to our better understanding of the
LKFT,~\cite{Jia:2016wyu,Ahmadiniaz:2016qqo,Ahmadiniaz:2015kfq,Fernandez-Rangel:2016zac,Bashir:2008ej,Bashir:2005wt,Bashir:2004yt,Bashir:2004hh,Bashir:2002sp,Bashir:2000iq}.

The need for the same in QCD was realized in the work by Bloch,
who constructs a model truncation which preserves MR, and
reproduces the correct leading order perturbative behavior through
assuming non-trivial cancellations involving the full quark-gluon
vertex in the quark self-energy loop,~\cite{Bloch:2002eq}.

Note that the quark propagator beyond ${\cal O}(\alpha)$ involves
gluon self interactions. These interactions introduce the color
factor $C_A$ in the adjoint representation. The same is true for
the transverse part of the quark-gluon vertex. In this section, we
provide these transverse form factors for the asymptotic limit,
$k^2 \gg p^2 \gg m^2$. For the Abelian part we have:
\bea \nn k^4\frac{\tau_1^{a}}{m}&=&C_a
\left(4-\xi\right)\ln\left(\frac{p^2}{k^2}\right) \,,
\\ \nn k^4\tau_2^{a}&=&\frac{C_a}{3}
\left(2\xi-1\right)\ln\left(\frac{p^2}{k^2}\right) \,,
\\ \nn
 k^2\tau_3^{a}&=&\frac{C_a}{3}
 \left(2-\xi\right)\ln\left(\frac{p^2}{k^2}\right) \,,
\\ \nn k^4\frac{\tau_4^{a}}{m}&=&-\frac{C_a}{12} \xi
\ln\left(\frac{p^2}{k^2}\right) \,, \nn \\
\nn k^2\frac{\tau_5^{a}}{m}&=&-\frac{C_a}{6} \xi
\ln\left(\frac{p^2}{k^2}\right) \,, \\
 \nn k^2\tau_6^{a}&=&\frac{C_a}{6}
\left(1+\xi\right)\ln\left(\frac{p^2}{k^2}\right) \,, \\
 \nn k^4\frac{\tau_7^{a}}{m}&=&\frac{C_a}{6} \xi
\ln\left(\frac{p^2}{k^2}\right) \,,
 \eea
 \bea
 \nn k^2\tau_8^{a}&=&C_a \ln\left(\frac{p^2}{k^2}\right) \,.
\eea For the non-Abelian part, we find:
\bea \nn k^4\frac{\tau_1^{b}}{m}&=&\frac{C_b}{12}
\left[18-5\xi-\xi^2\right]\ln\left(\frac{p^2}{k^2}\right) \,,
\\ \nn
 k^4\tau_2^{b}&=&\frac{C_b}{24}
\left[-2 + 7 \xi + \xi^2\right] \ln\left(\frac{p^2}{k^2}\right)
\,, \\  \nn k^2\tau_3^{b}&=&\frac{C_b}{24} \left(1-\xi\right)
\left( 4 -\xi \right) \ln\left(\frac{p^2}{k^2}\right)  \,,
\\ \nn
k^4\frac{\tau_4^{b}}{m}&=&\frac{C_b}{48} \xi
\left(3-\xi\right)\ln\left(\frac{p^2}{k^2}\right) \,,
\\ \nn
k^2\frac{\tau_5^{b}}{m}&=&\frac{C_b}{24}
\left[36-17\xi+3\xi^2\right]\ln\left(\frac{p^2}{k^2}\right) \,,
\\ \nn
 k^2\tau_6^{b}&=&\frac{C_b}{48}
\left[2-\xi+2\xi^2\right]\ln\left(\frac{p^2}{k^2}\right) \,,
\\ \nn
k^4\frac{\tau_7^{b}}{m}&=&-\frac{C_b}{24} \xi \left(1+\xi
\right)\ln\left(\frac{p^2}{k^2}\right) \,,
\\
k^2\tau_8^{b}&=&-\frac{C_b}{8} \left[6-5\xi+\xi^2
\right]\ln\left(\frac{p^2}{k^2}\right) \,. \eea
 In this asymptotic
limit, the leading structures in the massless quark-gluon vertex
are those proportional to $\tau_3$ and $\tau_6$, whose
corresponding basis vectors are proportional:
\begin{eqnarray}
T_{\mu}^{3 \, asy} = -T_{\mu}^{6 \, asy}= k^2 \gamma_{\mu} -
\not\!{k} k_{\mu}\equiv T_{\mu} \,, \label{Tau3YTau6Asy}
\end{eqnarray}
thus revealing the linear dependence between them. Naturally, the
leading behavior of the massless transverse vertex in this limit
thus reads as
\begin{eqnarray}
&& \hspace{-1cm}  \Gamma^{\mu}_T(p,k,q) \stackrel{{\tiny{k^2 \gg
p^2}}}{=}
(\tau_3-\tau_6) T^{\mu} \nn \\
&& \hspace{0.8cm} = -\alpha \frac{C_A(2-\xi) - 8 C_F (1-\xi)}{64
k^2 \pi} \ln \frac{p^2}{k^2} \, T^{\mu} \,.
\end{eqnarray}

\begin{figure}[h!]
\includegraphics[angle=0,width=0.5\textwidth]{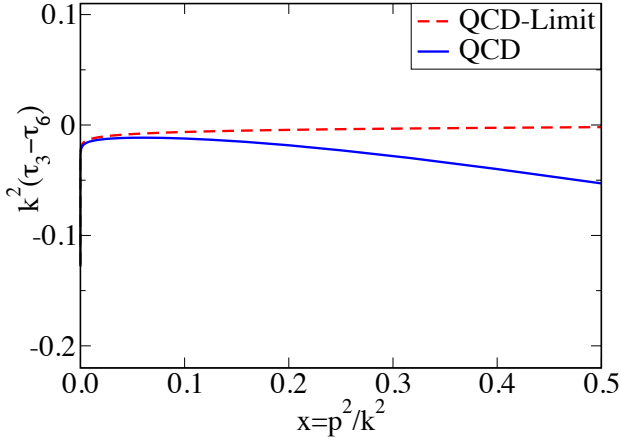}
\vspace{-0.5cm} \caption{The dimensionless combination $k^2 (
\tau_3 - \tau_6)$ of the transverse vertex. The solid (blue) line
is the numerical evaluation of the one-loop result. The dashed
(red) curve is the asymptotic result, valid for $x \rightarrow 0$
alone. As expected, the numerical result converges onto the
asymptotic analytically obtained value for $x \rightarrow 0$.}
\label{fig10}
\end{figure}

We confirm this result numerically in Fig.~(\ref{fig10}). This is
the QCD generalization of the QED result already derived in
~\cite{Curtis:1990zs}:
\begin{eqnarray}
&& \hspace{-1cm}  \Gamma^{\mu}_T(p,k,q) \stackrel{{\tiny{k^2 \gg
p^2}}}{=}
(\tau_3-\tau_6) T^{\mu} \nn \\
&& \hspace{1.04cm} =  \frac{ \alpha (1-\xi)}{8 k^2 \pi} \ln
\frac{p^2}{k^2} \, T^{\mu}  \,.
\end{eqnarray}
 We can carry out a similar analysis for the massive part.
The leading contribution in this limit comes from $\tau_{4,5,7}$,
whereas, $\tau_{1}$ chips in with a sub-leading term. Moreover, it
is worth noting that the following tensors are all proportional to
each other:
\begin{eqnarray}
&& T_{\mu}^{4 \, asy} = -k^2 T_{\mu}^{5 \, asy} = -2 T_{\mu}^{7 \,
asy} = k^2 (\gamma_{\mu} \not\!{k} - k_{\mu}) \nn \\
&=& (k^2 \gamma_{\mu} - k_{\mu} \not\!{k} ) \not\!{k} = T^{\mu}
\not\!{k} \,.
\end{eqnarray}
One can readily see that the leading terms of $\tau_{4,5,7}$ add
up to cancel. Hence, for small mass $m$, it is still the massless
transverse part which is dominant.

\begin{table*}
\begin{tabular}{|c|c|c|c|c|c|}
\hline
Vertex & \hspace{1.2cm} Structure & $a_i$ & MR & $\nu$\\
\hline
Bare & $\gamma_\mu$ & -- & No &\\
%\hline
BC~\cite{Ball:1980ay} & $
\Gamma_{\mu}=\sum_{i=1}^4 \lambda_i \, L_{\mu} = \Gamma_{\mu}^{BC} $ & -- & No & \\
%\hline
CP~\cite{Curtis:1990zs} & $ \Gamma_{\mu}= \Gamma_{\mu}^{BC}+\tau_6 T_{\mu}^{6}$ & $a_2=a_3=a_8=0$, $a_6=\frac{1}{2}$ & Yes & $C_F \alpha\xi/(4\pi)$\\
%\hline
BBCR~\cite{Bashir:2011dp} & $ \Gamma_{\mu} =
\Gamma_{\mu}^{BC}+\sum_{i=2,3,6,8} \tau_i \, T_{\mu}^i$ \, &
$a_6=-\frac{1}{2}, a_2 + 2 (a_3+a_8) = -2$ & Yes &
Numerical \\
ABG~\cite{Aslam:2015nia} & $ \Gamma_{\mu}  =
\Gamma_{\mu}^{BC}+\sum_{i=2,3,6,8}
\tau_i \, T_{\mu}^i$ \, & $ \hspace{-2.4mm} a_6=+\frac{1}{2}, a_2 + 2 (a_3+a_8) = 0 $ & Yes & $C_F \alpha\xi/(4\pi)$\\
QCLRS~\cite{Qin:2013mta} & $ \Gamma_{\mu}  =
\Gamma_{\mu}^{BC}+\sum_{i=2,3,6,8} \tau_i \, T_{\mu}^i$ \, & $
\hspace{0.0mm} a_2=a_6=0, a_3=1/2, a_8=-1 $ & Yes & Numerical
 \\
 %BCPQR~\cite{Binosi:2016wcx} & $ \Gamma_{\mu} =
 %\Gamma_{\mu}^{BC}+\sum_{i=1,3,\hat{45},8} \tau_i \, T_{\mu}^i$ \,
 %& $a_{2,6,7}=0$, $\mid a_{1,3} \in [-1,1] , a_{\hat{45}} \in
 %[-7,5] ,a_8 \in [-5,1]\mid$ & Yes &
 %Numerical.\\
 \hline
\end{tabular}
\caption{We compare different vertex {\em Ans$\ddot{a}$tze} as
regards the LKFT for the massless quark propagator. The first
three columns define the vertex we consider. The letters
correspond to the names of the authors. The fourth column shows
whether the quark propagator is MR or not. The last column states
the exact exponent of the quark propagator to determine if the
vertex complies with the exact prediction of the LKFT for the
leading log series, namely $\nu={C_F \alpha\xi}/{(4\pi)}$.}
 \label{tab_struc}
\end{table*}
Recently, there has been a vertex {\em Ansatz} proposed in
~\cite{Binosi:2016wcx}. The form of the vertex is similar to the
ones used in Table~\ref{tab_struc}. However, the coefficients
$a_i$ depend explicitly on the angle between $k$ and $p$ and this
dependence, in some cases, continues to persist in the asymptotic
limit $k^2 \gg p^2$, preventing a direct similar comparison. The
masselss transverse part of the vertex is crucial in ensuring the
MR of the quark propagator, as was discussed in detail
in~\cite{Aslam:2015nia}. A popular choice of the vertex {\em
Ansatz} consists in proposing the following form:
 \begin{subequations}
  \label{ansatz}
  \bea
  \label{tau2}
  \tau_2(k^2,p^2) &=& \frac{a_2\, {\cal D}_F(k^2,p^2)}{(k^2+p^2)} \,,\\
  \label{tau3}
  \tau_3(k^2,p^2) &=& a_3\, {\cal D}_F(k^2,p^2)\,, \\
  \label{tau6}
  \tau_6(k^2,p^2) &=& a_6 \, \frac{(k^2+p^2)}{(k^2-p^2)} \, {\cal D}_F(k^2,p^2) \,, \\
  \label{tau8}
  \tau_8(k^2,p^2) &=& a_8 {\cal D}_F(k^2,p^2)\,,
  \eea
  \end{subequations}
 where
 \bea
  {\cal D}_F(k^2,p^2) &=& \frac{1}{(k^2 - p^2)}
    \left[    \frac{1}{F(k^2)} - \frac{1}{F(p^2)} \right] \,.
    \nn
 \eea
  Note that ${\cal D}_F(k^2,p^2)$ starts at one loop perturbation theory and
  contains a multiplicative color factor $C_F$ at that level, as expected from
  the one loop calculation of the quark propagator. Based upon the
  choice of the $a_i$, we make contact with different choices of
  the quark-gluon vertex adopted in the literature. With such a
  choice of Abelian-type vertex in QCD, different choices for
  $a_i$ determine whether a MR solution is possible and if it
  correctly reproduces leading logarithm behavior to all orders for the
  quark wavefunction renormalization, as dictated by the
  generalized LKFT for QCD introduced in~\cite{Aslam:2015nia},
  namely, $F(p^2) \propto (p^2)^{\nu=C_F \alpha \xi/(4 \pi)}$. We can
  compare and contrast different vertex {\em Ans$\ddot{\rm a}$tze}, as explained
  in Table~(\ref{tab_struc}), to see if they permit a MR solution
  and if the resulting anomalous dimension is $\nu=C_F \alpha \xi/(4 \pi)$.
  The contribution of $C_A$ begins at the next level in
  perturbation theory.

We now move onto discussing the on-shell limit in the next
section.

%%%%%%%%%%%%%%%%%%%%%%%%%%%%%%%%%%%%%%%%%%%%%%%%%%%%%%%%%%%%%%%%%
%%%%%%%%%%%%%%%%%%%%%%%%%%%%%%%%%%%%%%%%%%%%%%%%%%%%%%%%%%%%%%%%%
\section{The On-Shell limit}
\label{sec:OnShellLimit}
%%%%%%%%%%%%%%%%%%%%%%%%%%%%%%%%%%%%%%%%%%%%%%%%%%%%%%%%%%%%%%%%%
%%%%%%%%%%%%%%%%%%%%%%%%%%%%%%%%%%%%%%%%%%%%%%%%%%%%%%%%%%%%%%%%%

In this section we present some ``physically" relevant results for
the on-shell limit $p^2=k^2=m^2$ and $q^2=0$. The Dirac and Pauli
form factors, $F_1(q^2)$ and $F_2(q^2)$, respectively, define the
Gordon decomposition of the quark current as follows:
\begin{eqnarray}
\overline{u}(p) \Gamma_{\mu}(p,k,q)\Big|_{k^2=p^2=m^2} u(k) &=&
\nonumber \\
&& \hspace{-4.5cm} \overline{u}(p) \Bigg\{ F_1(q^2) \gamma_{\mu} -
\frac{F_2 (q^2)}{2m} \; \sigma_{\mu \nu} q^{\nu} \Bigg\} u(k) \,,
\nonumber
\end{eqnarray}
where the spinors, $\overline{u}(p)$ and $u(k)$, satisfy the Dirac
equation:
\begin{eqnarray}
\overline{u}(p) \not\!{p} &=& m \, \overline{u}(p) \,, \nonumber \\
\not\!{k} \, u(k) &=& m \, u(k) \,. \nonumber
\end{eqnarray}
The anomalous chromomagnetic moment (ACM) of quarks can be
identified as $F_2(q^2)$ for $q^2 \rightarrow 0$. The Abelian
version of this decomposition with $C_F=1$ and $C_A=0$ is the
electron-photon vertex of quantum electrodynamics. The great
successes of the Dirac equation is the prediction of the magnetic
moment of a charged fermion  ${\bm {\mu}} = {eg}/{(2 m)} {\bm S}$.
The radiative corrections lead to~\cite{Schwinger:1948iu}
 \bea
 \frac{e}{2 m} \Rightarrow \left( 1 + \frac{\alpha}{2 \pi} \right) \frac{e}{2
 m}.
 \eea
These corrections are now known to a much higher order in
perturbation theory~\cite{Aoyama:2011dy}.

 Note that the quark-gluon vertex differs from the electron-photon vertex
already at one loop, by the contributions of an additional Feynman
diagram, involving the triple-gluon vertex. In fact, apart from
introducing additional color structure, this non-Abelian diagram
introduces, at the one-loop level, a kinematical structure which
is absent in the QED.

\begin{figure}[h!]
\centering
\includegraphics[angle=-90,width=0.5\textwidth]{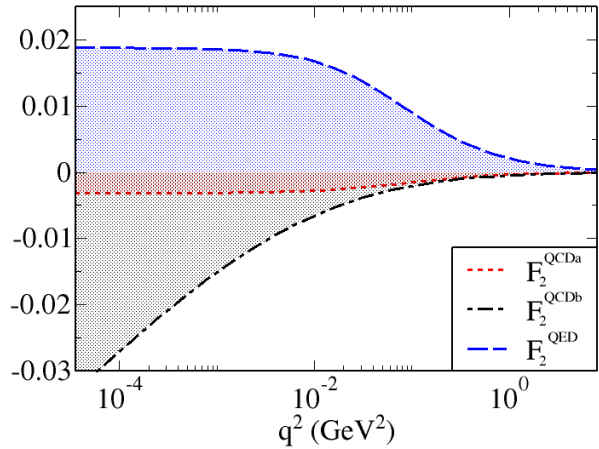}
\caption{$F_{2}(q^{2})$ in the on-shell case.} \label{fig11}
\end{figure}

One is naturally tempted to calculate $F_2(q^2)$. It can be
expressed in terms of the quark-gluon vertex form factors as
follows:
\begin{eqnarray}
F_2 (q^2) &=& - 2m  \lambda_{2}^{\rm os}(q^2) + \lambda_{3}^{\rm
os}(q^2)
- \frac{1}{2} q^2 \tau_{1}^{\rm os}(q^2) \nonumber \\
&& -m q^2 \tau_{2}^{\rm os}(q^2) - \tau_{5}^{\rm os}(q^2) +
\frac{1}{2}
q^2 \tau_{7}^{\rm os}(q^2) \nonumber \\
&& - m \tau_{8}^{\rm os}(q^2) \,,
\end{eqnarray}
where we have introduced as a simplifying notation
$\lambda_{i}^{os},\tau_{i}^{os} (q^2) \equiv \lambda_{i},\tau_{i}
(m^2,m^2,q^2)$. At one-loop perturbation theory, in Landau gauge,
the Abelian and non-Abelian contributions for the ACM are plotted
in Fig.~(\ref{fig11}), as a function of gluon momenta $q^2$, for a
current quark mass $m=.115$ GeV, and $\alpha=.118$. These
contributions can be analytically expressed as
\begin{eqnarray}
F_{2}^{a}(q^2) &=& -\frac{8 C_a m^2}{(q^2-4 m^2)} f(q^2) \,, \nonumber \\
F_{2}^{b}(q^2) &=& \frac{2 C_b m^2}{(q^2-4m^2)} \Bigg\{ 8m^2 -2q^2
 - 6m^2 q^2 \varphi_{1}^{os} (q^2) \nonumber \\
&& \hspace{1.5cm} +(8m^2+q^2) \ln \left( -\frac{q^2}{m^2} \right)
\Bigg\} \,,
\end{eqnarray}
where we define $\varphi_{1}^{os} (q^2) \equiv
\varphi_{1}(m^2,m^2,q^2)$.

\vspace{0.5cm}

It is straightforward to see that for the soft gluon limit,
$q^2=0$, the Abelian contribution for the ACM reduces to the
non-Abelian counterpart of Schwinger's result, $F_{2}^{a}(0) = -
\alpha /12 \pi$, already derived in~\cite{Chang:2010hb}. On the
other hand, the corresponding non-Abelian contribution vanishes
for a massless current quark, $m=0$, as reported in the same
article. However, it yields a divergence,\cite{Choudhury:2014lna},
for a non-zero quark mass, $m \neq 0$. We find this divergence to
be logarithmic. For deep infrared gluon momenta it behaves as
$F_2^{b}(q^2 \rightarrow 0)= C_b \ln \left( - q^2/m^2 \right)$. Of
course, perturbation theory in QCD is not the way to explore deep
infrared region. All perturbative conclusions will be taken over
by non-perturbative effects, overshadowing this divergence.
\\

%%%%%%%%%%%%%%%%%%%%%%%%%%%%%%%%%%%%%%%%%%%%%%%%%%%%%%%%%%%%%%%%%
%%%%%%%%%%%%%%%%%%%%%%%%%%%%%%%%%%%%%%%%%%%%%%%%%%%%%%%%%%%%%%%%%
\section{Conclusions}
\label{sec:Conclusions}
%%%%%%%%%%%%%%%%%%%%%%%%%%%%%%%%%%%%%%%%%%%%%%%%%%%%%%%%%%%%%%%%%
%%%%%%%%%%%%%%%%%%%%%%%%%%%%%%%%%%%%%%%%%%%%%%%%%%%%%%%%%%%%%%%%%

In this paper, we give a detailed numerical analysis of all the
form factors defining the quark-gluon vertex at the one-loop level
in different kinematical limits of interest: symmetric, asymptotic
and on-shell. The symmetric limit of momenta is rather
well-behaved in the infrared region, where all the Abelian form
factors converge to finite values. The non-Abelian form factors
are only logarithmically divergent. Most noticeably, all the
longitudinal form factors are infrared finite. Any
non-perturbative construction of these form factors or their
computation on the lattice must comply with this requirement. The
on-shell limit enables us to compute anomalous magnetic moment of
quarks and confirm our numerical computation with the
corresponding results known for QED and
QCD,~\cite{Schwinger:1948iu,Chang:2010hb}. The triple gluon
contribution to the ACM of quarks is logarithmically divergent. We
find exact analytical expression for this divergence. The
asymptotic results have implications for the multiplicative
renormalizability of the fermion propagator both in QED and QCD.
This connection is exposed through the LKFT, allowing us to
analyze various {\em Ans$\ddot{a}$tze} put forward in the
literature. Our study provides us with quantitatively detailed
results for kinematical limits of interest and hence a guideline
to all non-perturbative constructions of the
corresponding vertices as well as the lattice computations. \\

\noindent {\bf{Acknowledgements}} Research supported by: CIC
(UMSNH) and CONACyT Grant nos. 4.10 and CB-2014-22117.

%%%%%%%%%%%%%%%%%%%%%%%%%%%%%%%%%%%%%%%%%%%%%%%%%%%%%%%%%%%%%%%%%
%%%%%%%%%%%%%%%%%%%%%%%%%%%%%%%%%%%%%%%%%%%%%%%%%%%%%%%%%%%%%%%%%
\bibliography{SDE-References}

\end{document}